\documentclass[aps,prl,twocolumn,superscriptaddress,nobibnotes,nofootinbib]{revtex4-1}

\usepackage{bbm}
\usepackage{mathrsfs}
\usepackage{amsmath}
\usepackage{amsfonts}
\usepackage{graphicx}
\usepackage{subfigure}
\usepackage{dcolumn}
\usepackage{bm}
\usepackage{color}
\usepackage{natbib}
\usepackage{amssymb}
\usepackage{xcolor}
\usepackage{physics}
\usepackage[normalem]{ulem}
\usepackage{lipsum}
\usepackage[colorlinks=true,
	        urlcolor=blue,
	        citecolor=blue,
	        linkcolor=blue
            ]{hyperref}

\begin{document}

\title{Emergence of drifted diffusion in quantum walks with subspace restart}

\author{Liwei Qiao}
\affiliation{School of Physics and Key Laboratory of Quantum State Construction and Manipulation (Ministry of Education), Renmin University of China, Beijing 100872, China}
\author{Ruoyu Yin}
\email{Contact author:yin.ruoyu.5v@kyoto-u.ac.jp}
\affiliation{School of Physics, Peking University, 100871 Beijing, China}
\affiliation{Department of Physics \#1, Graduate School of Science, Kyoto University, Kyoto 606-8502, Japan}

\author{Wei Zhang}
\email{Contact author:wzhangl@ruc.edu.cn}

\affiliation{School of Physics and Key Laboratory of Quantum State Construction and Manipulation (Ministry of Education), Renmin University of China, Beijing 100872, China}
\affiliation{Beijing Academy of Quantum Information Sciences, Beijing 100193, China}

\begin{abstract}
Restart of a quantum process is typically modeled as a global reinitialization that erases the system's entire history.
Here we introduce subspace restart,
a protocol that periodically resets only the internal degrees of freedom while preserving the spatial distribution,
as a tunable knob for the quantum-to-classical crossover.
Using the discrete-time quantum walk as an example, 
we show that this selective reset drives the walker into an {\em engineered} drifted-diffusion regime.
This phenomenon can be understood by a Huygens-Fresnel mechanism,
where each restart fragments the wave function into a set of independent secondary sources to
screen long-range correlations and isolate a robust classical backbone,
whose drift and diffusivity are set by the geometric orientation of the initial coin and the restart period.
Residual quantum interference, confined to effective light cones,
survives only as a short-range correction
that renormalizes these coefficients 
and imprints periodic modulations on the cumulants.
Our results establish subspace restart as a route to controlling the quantum-to-classical crossover in synthetic lattices.
\end{abstract}
\maketitle

\textit{Introduction}.---
Quantum walk provides a versatile framework for studying coherent transport, interference-driven spreading, and quantum-walk-based algorithms \cite{Kempe2003,Shenvi2003,Childs2003,Ambainis2007,Childs2009,MULKEN2011,VenegasAndraca2012,Portugal2013}. 
Unlike classical random walks, whose dynamics are governed by stochastic transition probabilities and typically lead to diffusive spreading on regular lattices
\cite{Montroll1965,Hughes1996random}, 
quantum walks evolve through probability amplitudes,
allowing superposition and interference to fundamentally reshape transport behavior \cite{PhysRevA.48.1687,Gutmann1998,Ambainis2001,Mackay2002}. 
Among different formulations, discrete-time quantum walks (DTQWs) are particularly appealing because their internal coin degree of freedom offers a tunable mechanism for controlling spatial propagation \cite{Tregenna2003,PhysRevA.77.032326,PhysRevLett.104.050502}.
In ideal coherent settings, such interference effects can give rise to ballistic spreading and reveal transport features with no classical analog \cite{nayak2000quantumwalkline,konno2002quantum,Perets2008,Karski2009}. 
At the same time, these coherence-dependent phenomena are intrinsically sensitive to non-unitary processes. Environmental dephasing, controlled decoherence, and intermediate measurements can suppress interference and gradually drive the dynamics toward classical-like diffusion \cite{Brun2003,KendonTregenna2003,ROMANELLI2005,Oliveira2006,Kendon2007,Maloyer2007,Annabestani2010,Broome2010,Schreiber2011,Alberti2014}. 
Consequently, DTQW serves as a highly controllable platform for investigating the interplay between coherent transport and decoherence, and the crossover between quantum classical dynamics \cite{Kendon2007,Maloyer2007,Annabestani2010,Broome2010,Schreiber2011,Alberti2014,Eliahu2021,Bressanini2022,Kua2025}.

Beyond such coherence-degrading processes,
resetting protocols offer a complementary non-unitary mechanism for controlling quantum dynamics
\cite{PhysRevB.98.104309,PhysRevB.104.L180302,Majumdar2023entanglement,wald2025stochastic,Puente2024quantumstate,rbtb-8d27,gotta2026scars,murauer2026,carollo2026nonMarkovianity,Jafari2026}. 
By interrupting system evolution at prescribed or random intervals and reinitializing it according to a given rule,
resetting mechanisms have been widely used 
to manipulate relaxation, stationary behavior, first-passage properties, and monitored dynamics 
across both classical and quantum domains \cite{evans2011diffusion,pal2016diffusion,chechkin2018random,evans2020stochastic,pal2017first,reuveni2016optimal,PhysRevB.98.104309,PhysRevB.104.L180302,Majumdar2023entanglement,Liu2023Semi,Puente2024quantumstate,rbtb-8d27,Kumar2025,yin2025restart,wald2025stochastic,10.1063/5.0261830,biswas2026threshold,gotta2026scars,murauer2026,carollo2026nonMarkovianity,Jafari2026}. 
In most of existing protocols, reset acts as a {global} renewal
\cite{PhysRevB.98.104309,PhysRevB.104.L180302,Majumdar2023entanglement,wald2025stochastic,rbtb-8d27,li2020photonic,PhysRevE.111.044143,Yin2025resonance,solanki2025relaxation,gotta2026scars,murauer2026,carollo2026nonMarkovianity,Jafari2026}, under which
the {\em entire} quantum state is returned at once to a single configuration, 
typically its initial condition to realize a restart. 
While effective for relaxation and search algorithms \cite{evans2011diffusion,PhysRevLett.130.050802,PhysRevResearch.7.023069,PhysRevE.110.034132,Bao2025,solanki2025relaxation},
this drastic intervention obliterates the system's accrued history and quantum correlations, 
effectively treating the quantum evolution as a black box to be rebooted. 
It is therefore natural to ask whether one can design a gentle resetting protocol that partially refreshes the dynamics not by erasing the entire history but by harnessing it,
and in doing so,
whether resetting can be turned into a precise knob for the quantum-to-classical crossover.

Here, we introduce {\em subspace restart},
a protocol that periodically resets only part of the degrees of freedom to their initial state, while leaving the others unperturbred. 
For DTQW model, this scheme avoids the limitations of global resetting, 
and differs from recently developed subsystem resetting, 
which reinitializes a separable part of a {classical} many-body system 
\cite{Gupta2024,Gupta2025,majumder2026analytical}. 
Specifically, we consider restart acting only on the internal degrees of freedom (coin) but not the spatial probability distribution. This allows the walker to retain a memory of its location
while refreshing its spreading capability. 
We demonstrate that such protocol drives the quantum walker into a regime of engineered drifted diffusion, which bridges between unitary quantum dynamics and classical stochastic transport and features highly tunable drift and diffusivity.
To understand this phenomenon, we introduce a Huygens-Fresnel-type decomposition mechanism. 
where the reset fragments the wave function into a set of independent secondary sources such that the 
incoherent sum forms a robust classical backbone with diffusivity scaling linearly in the restart period
and drift fixed by the geometric orientation of the initial coin state.
Residual quantum coherence survives only within an
{effective light cone} of each source,
contributing a short-range interference correction 
that renormalizes these bare coefficients without altering their scaling.
Our results demonstrate the ability of subspace restart to tune the crossover between quantum and classical transport, and can be readily tested in existing quantum walk experimental systems \cite{carolan2015universal,nitsche2016quantum,goel2025quantuminformationprocessingspatially}.

\begin{figure}[tp]
    \centering
    \includegraphics[width=1.0\linewidth]{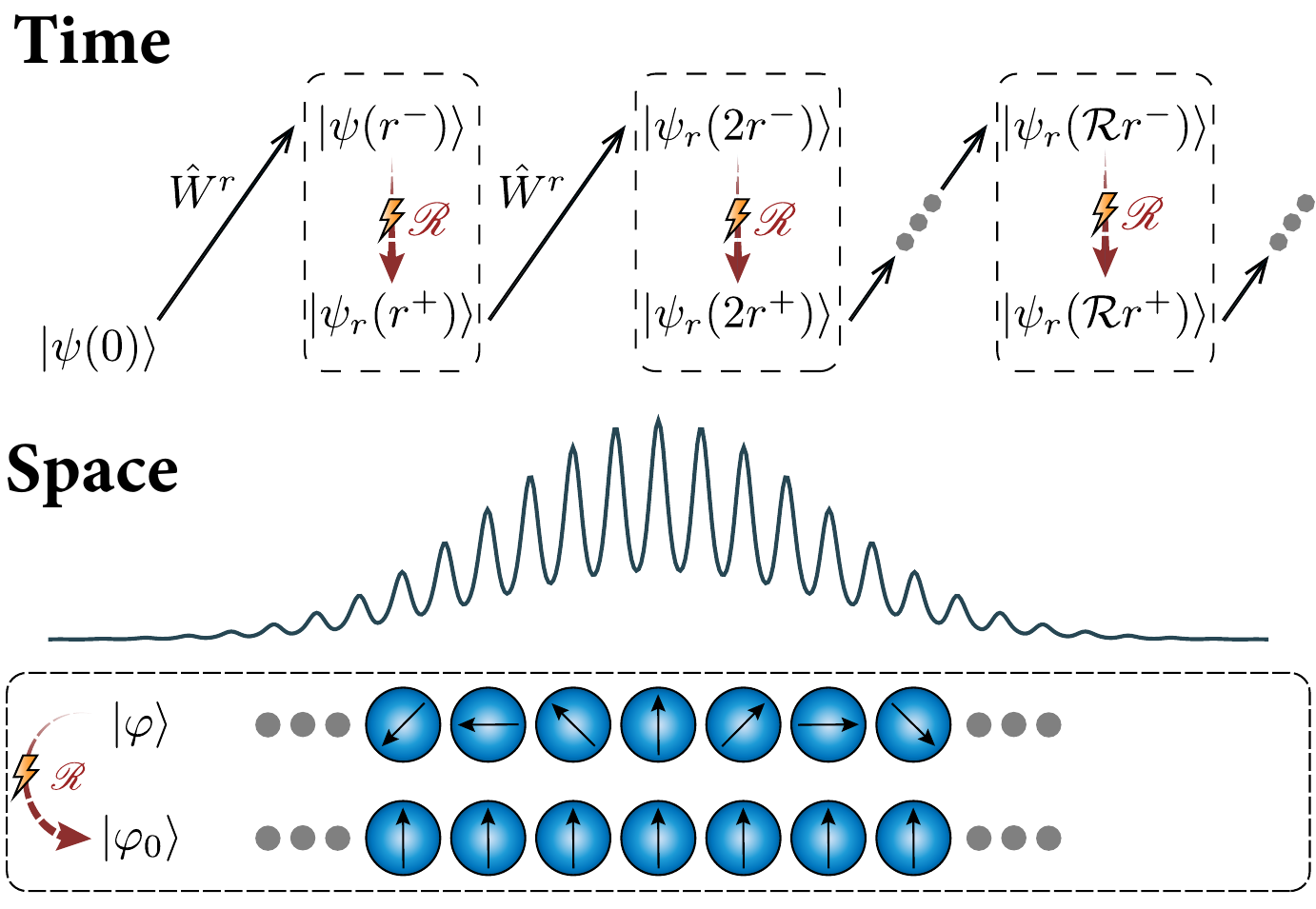}
    \caption{
    Schematic of a discrete-time quantum walk with periodic subspace restart. 
    (Top) Temporal evolution: 
    uninterrupted unitary dynamics $\hat{W}^{r}$ are interleaved with instantaneous coin resets $\mathscr{R}$ (dashed boxes).
    (Bottom) Action of $\mathscr{R}$ in real space: 
    the spatial probability profile is left invariant, 
    while the internal coin state at every lattice site (blue spheres) is reset from the evolved $|\varphi\rangle$ back to the initial $|\varphi_{0}\rangle$.}
    \label{model}
\end{figure}

{\em Model}.---
We consider a DTQW on an infinite one-dimensional (1D) lattice $\mathbb{Z}$,
augmented by a periodic restart mechanism.
The walker resides in the Hilbert space $\mathcal{H} = \mathcal{H}_{p} \otimes \mathcal{H}_{c}$, spanned by position states $\{\ket{x}\}$ and internal coin states $\{\ket{\uparrow}, \ket{\downarrow}\}$.
The dynamics are governed by the unitary step operator
$\hat{W} = \hat{S} (\mathbb{I}_p \otimes \hat{C})$,
where $\hat{C}$ is the coin toss and $\hat{S} = \sum_{x} ( \ket{ x+1 } \bra{ x } \otimes \ket{ \uparrow } \bra{ \uparrow } + \ket{ x-1 } \bra{ x } \otimes \ket{ \downarrow } \bra{ \downarrow } )$
is the coin-dependent shift 
\cite{PhysRevA.48.1687, nayak2000quantumwalkline}.
We focus on the standard Hadamard walk 
$\hat{C} = \hat{H} = (\hat{\sigma}_x + \hat{\sigma}_z)/\sqrt{2}$
with $\hat{\sigma}_x, \hat{\sigma}_z$ the Pauli matrices,
initialized at the origin $\ket{\psi(0)} = \ket{0} \otimes \ket{\varphi_0}$
with a generic coin state
$\ket{ \varphi_0 } = \cos(\theta/2)\ket{ \uparrow } + e^{i\phi} \sin(\theta/2) \ket{ \downarrow }$.
In the absence of restart,
the state at time $n$ is given by 
\begin{equation}\label{eq2_freeEvolution}
    \begin{aligned}
        \ket{ \psi(n) } 
        =   \hat{W}^{n} \ket{ \psi(0) } 
        =   \sum_{x,s} 
            \alpha_{s}(x,n|{0})\, { \ket{ x,s }}, 
    \end{aligned}
\end{equation}
where $\alpha_{s}(x,n|0)$ is the amplitude for the walker
starting at $0$ to reach site $x$ with spin $s$.
The corresponding probability distribution of spatial occupation
$P(x,n) = \sum_s |\alpha_s(x,n|0)|^2$
exhibits the well-known ballistic spreading  
with distinct drift velocities determined by the initial angles $(\theta, \phi)$ \cite{nayak2000quantumwalkline}.

{\em Subspace restart and the source picture.}---
We introduce a periodic restart operation $\mathscr{R}$,
applied every $r$ steps,
which resets the internal degree of freedom
while preserving the spatial distribution,
as illustrated in Fig.~\ref{model}.
Physically, this corresponds to disentangling the coin from the position \cite{Kendon2007,Maloyer2007}.
Mathematically,
the map $\ket{\psi} \to \mathscr{R}(\ket{\psi})$ projects the system onto the initial coin state $\ket{\varphi_0}$ at every site,
\begin{equation}
    \begin{aligned}
    \ket{\psi_{r}(r)}
    = \mathscr{R} \big( \ket{ \psi(r) } \big)
    = \sum_{x} \sqrt{P(x,r)} \ket{ x } \otimes \ket{ \varphi_0 } .
    \end{aligned}
\end{equation}
Here the subscript $r$ denotes the restart period.
Note that the post-reset amplitude is $\sqrt{P(x,r)}$,
ensuring probability conservation;
this reflects the loss of local relative phases between spin components (decoherence) while retaining the population density.

\begin{figure}[htbp]
    \centering
    \includegraphics[width=1.0\linewidth]{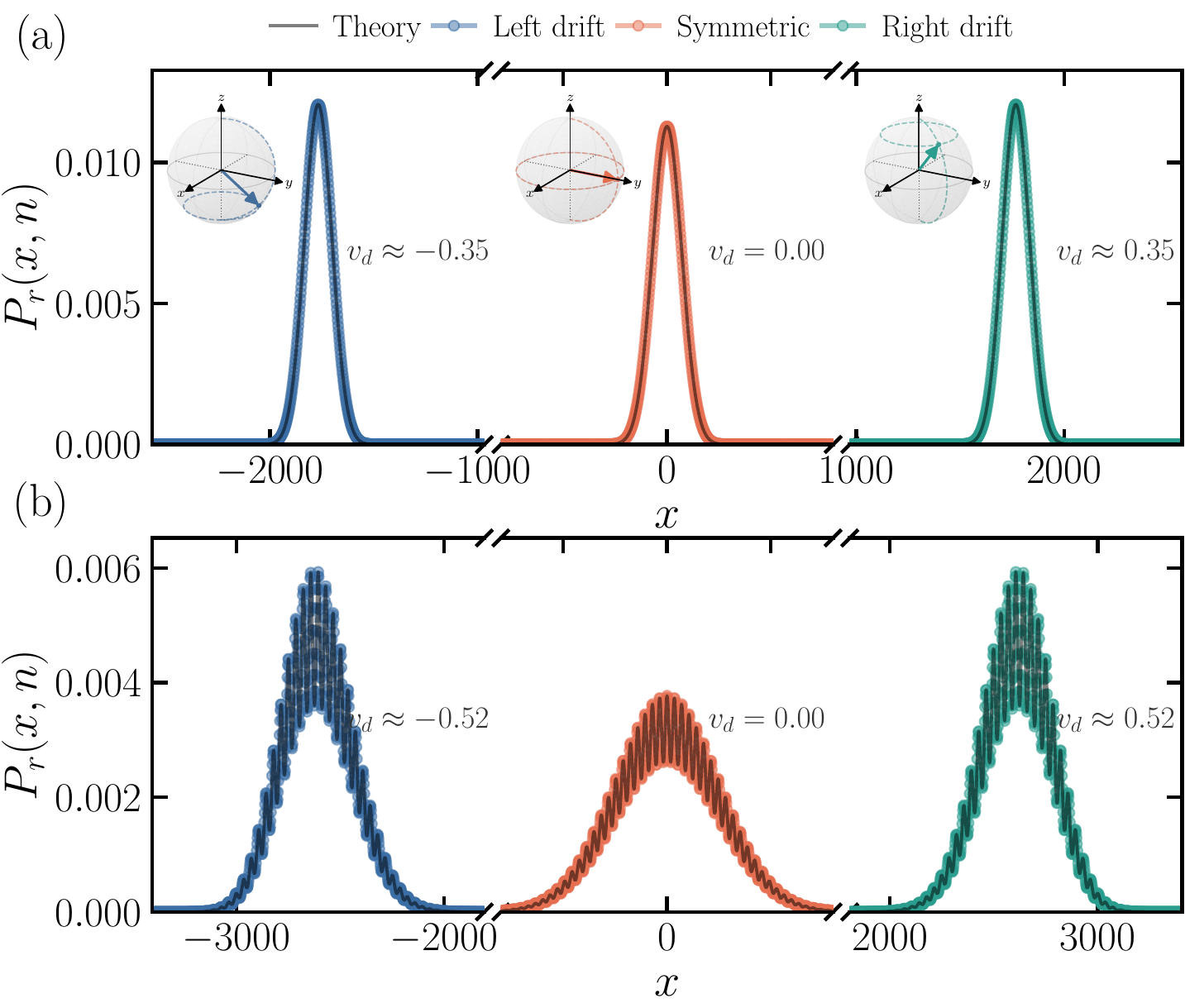}
    \caption{Occupation probability $P_{r}(x, n)$ at $n=5000$ steps for restart periods 
    (a) $r=1$ and (b) $r=25$,
    with three coin orientations (Bloch spheres): 
    left drift $(\theta,\phi)=(3\pi/4, 2\pi/3)$ (blue), 
    symmetric spreading $(\pi/2,\pi/2)$ (orange), and right drift $(\pi/4,\pi/3)$ (green).
    At $r=1$ the distribution is a clean Gaussian,
    while at $r=25$ a Gaussian envelope persists but is dressed by short-range interference fringes, signaling that the macroscopic transport classicalizes 
    while local quantum coherence survives within each cycle.
    Without specification, only lattice sites with nonzero probability are shown.
    }
    \label{coherence}
\end{figure}

The reset generates a source picture.
Since every populated site is reinitialized in the same coin state $\ket{\varphi_0}$,
translational invariance turns the distribution at $n=\mathcal{R}r$ into identical secondary sources.
For $n = \mathcal{R} r + \tilde{n}$,
with $\mathcal{R}\ge0$ the number of restart cycles and $\tilde{n} \in (0, r]$  the residual time,
the state is
\begin{equation}
    \ket{ \psi_{r}(n) }
    = \sum_{x,x^\prime,s} \sqrt{P_{r}(x,\mathcal{R} r)} \, \alpha_{s}(x^\prime,\tilde{n}|0) \ket{ x+x^\prime, s }.
    \label{Wave source}
\end{equation}
Each source therefore emits a translated copy of the same wavelet
$\alpha_s(x^\prime,\tilde n|0)$, 
giving a Huygens-Fresnel-type construction. 
Their coherent interference determines
\begin{equation}
    P_{r}(x,n) = \sum_{s} 
    \bigg|
    \sum_{x^\prime} 
    \sqrt{P_{r}(x^\prime,\mathcal{R}r)} 
    \,
    \alpha_s (x - x^\prime,\tilde{n}|0) 
    \bigg|^2.
    \label{iteration}
\end{equation}
Here $\sqrt{P_r}$ gives the source weights, while $\alpha_s$ carries the propagation phase.
Specifically,
$P_{r}(x,0)=\delta_{x,0}$,
and before the first restart,
$P_{r}(x,{n}\le r) = P(x,{n}) = \sum_s |\alpha_s(x,{n}|0)|^2$.
Equation~\eqref{iteration} thus iteratively captures the competition between ballistic spreading within each cycle and periodic removal of coin-position entanglement,
in exact agreement with direct evolution under $\hat W$ and $\mathscr R$  demonstrated in Fig.~\ref{coherence}.
However, the result also reveals a striking feature:
the complex interference fringes typical of quantum walks gradually give way to a classical-like Gaussian envelope,
suggesting that subspace restart suppresses long-range coherence.
We next explore this emergence of classicality by examining limiting regimes of the restart period $r$. 

{\em The classical random walk limit.}---
To identify the onset of classicality,
we isolate the dynamical component driven solely by incoherent spreading of probability,
explicitly neglecting the interference between wavelets.
This leads to the diagonal approximation,
which anchors the dynamics to the incoherent sum of individual source intensities,
\begin{equation}\label{convolution}
    \begin{aligned}
    P_{r}^{\rm inc}\left( x,n \right)
    = \sum_{x^\prime} P_{r}^{\rm inc}(x^\prime,r\mathcal{R})  \sum_{s}|\alpha_s (x - x^\prime,n-r{\cal R}|0)|^2.
    \end{aligned}
\end{equation}
While the validity of this expression is limited by the interference of overlapping waves,
necessitating corrections discussed later,
it crucially isolates a robust {classical backbone}.
This discrete convolution structure captures the essential incoherent transport governing the system. 
The central limit theorem dictates that this incoherent contribution must asymptotically relax toward a Gaussian distribution, regardless of the restart interval $r$
\cite{feller1971introduction}. 
The exact distribution $P_{r}(x,n)$ can thus be understood as this smooth Gaussian core, dressed by perturbative corrections arising from residual inter-source interference.

This approximation is naturally motivated by the limit of stepwise restarts, $r=1$.
In this regime, the rapid memory erasure arrests the ballistic propagation
before the wavelets can expand sufficiently to overlap and interfere.
The waves remain spatially distinguishable,
leading to a perfect cancellation of off-diagonal coherences.
Consequently, the diagonal approximation becomes exact, i.e. $P_{1}(x,n)= P^{\rm inc}_{1}(x,n)$. 
Substituting the one-step amplitudes obtained with Eq.~\eqref{eq2_freeEvolution} into Eq.~\eqref{convolution} or \eqref{iteration}, we reduce the dynamics to a classical master equation,
\begin{equation}\label{classicalization}
    P_{1}(x,n+1) 
    = w_+ P_{1}(x-1, n) + w_- P_{1}(x+1, n),
\end{equation}
with transition probabilities
$w_\pm = (1\pm\sin{\theta}\cos{\phi})/2$
determined by the orientation of the initial coin state on the Bloch sphere ~\cite{supp}.
This recovers a classical random walk \cite{Redner_2001}, confirming that in the high-frequency restart limit, the quantum walk is stripped of its coherence, exposing the Gaussian classical core, as witnessed in Fig.~\ref{coherence}(a).
The drift velocity is $v_d=w_+-w_- =\sin{\theta}\cos{\phi}$, and the diffusion coefficient is $D=(1-v_d^2)/2=2w_+w_-$,
determining the shape of the resulting Gaussian,
i.e. $P_G(x,n)\sim \exp\left[ -(x - n v_d)^2 /4 n D \right]$.
Later we will return to these parameters.
Notably, a symmetric random walk ($v_d=0$) arises strictly
when the initial coin state lies in the $Y$-$Z$ plane (zero $x$-projection).
Since the Hadamard gate maps the $X$-axis to the $Z$-axis ($\hat{H} \hat{\sigma}_{x} \hat{H} = \hat{\sigma}_{z}$ and vice versa),
any initial state with vanishing $\langle \hat{\sigma}_x \rangle$ is rotated onto the Bloch sphere's equator ($\langle \hat{\sigma}_z \rangle = 0$).
This yields an equal superposition of $\ket{\uparrow}$ and $\ket{\downarrow}$
after the coin flip,
ensuring unbiased spatial propagation.

Therefore, tuning the coin parameters $\theta$ and $\phi$ offers precise control over transport, interpolating between symmetric diffusion ($\langle \hat{\sigma}_x \rangle_{\varphi_0}=0$) and directed drift ($\langle \hat{\sigma}_x \rangle_{\varphi_0}\neq0$). 
Crucially, the limit of $r=1$ uncovers a remarkable correspondence:
the internal coin asymmetry that dictates the ballistic group velocity in standard quantum walks directly translates into the stochastic drift of the emergent classical random walk.
We next demonstrate the generality of this connection.

{\it Emergence of drifted diffusion.}--- 
For a general $r$, we conjecture that the coin-controlled drift persists, 
now superimposed on a Gaussian envelope generated by the periodic reset's effective dephasing.
Although residual coherence sustains local interference oscillations, 
the macroscopic spread in Fig.~\ref{coherence}(b) is consistent with this Gaussian-like trend.
To quantify this,
we define the {Gaussian backbone}
from the first two cumulants,
\begin{equation}
\label{eq:gaussian_ansatz}
P_{\mathcal{G}}(x,n) = 
\frac{2}{\sigma_n \sqrt{2\pi}} 
\exp
\left[ 
-\frac{(x - \mu_n)^2}{2\sigma_n^2} 
\right].
\end{equation}
The prefactor $2$ accounts for the bipartite parity of the lattice support,
where the walker resides strictly on even (or odd) sublattices at any given step.
Here, $\mu_n := \langle x(n) \rangle$ represents the mean displacement.
To capture the total spatial excursion,
we utilize the mean squared displacement (MSD),
$\langle x^2(n) \rangle$.
The intrinsic diffusive spreading is then isolated by subtracting the drift component from the MSD,
yielding the variance $\sigma^2_n := \langle x^2(n) \rangle - \mu^2_n$.
In the long-time limit, this variance determines the effective diffusion coefficient via $\sigma^2_n \rightarrow 2 D_{\rm eff} n$.
The Gaussian backbone washes out the quantum fringes as visualized in Fig.~\ref{coherence}(b) and further discussed in the Supplemental Material~\cite{supp},
but isolates precisely the two cumulants that govern long-time transport.

\begin{figure}[tp]
    \centering
    \includegraphics[width=1.0\linewidth]{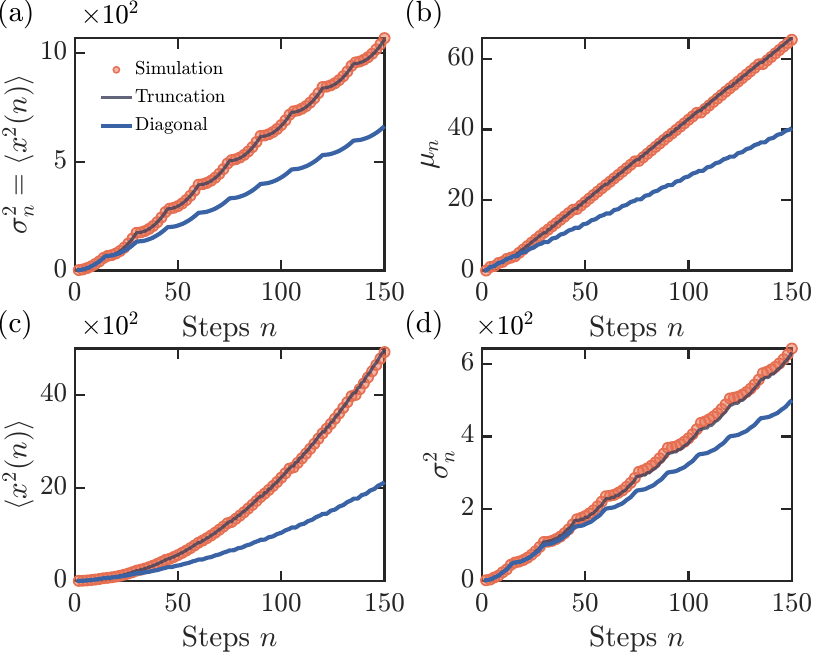}
    \caption{Accumulation of statistical quantities versus steps $n$ under subspace restart 
    with period $r=15$ and truncation $\Delta_{\max}=8$.
    (a) Variance $\sigma^2_n$ for the symmetric
    coin $(\theta,\phi)=(\pi/2,\pi/2)$.
    % coinciding with the MSD $\langle x^2(n) \rangle$.
    (b)--(d) Mean displacement $\mu_n$, MSD $\langle x^2(n) \rangle$ 
    and variance $\sigma^2_n$ for an asymmetric coin $(\theta,\phi)=(0,0)$. 
    Orange circles represent exact numerical simulations, 
    and the solid lines correspond to theory [Diagonal (blue), Truncation (black)]. 
    The diagonal ansatz captures the linear-in-$n$ scaling and the periodic staircase modulation, 
    while the truncated reconstruction with a small light-cone cutoff $\Delta_{\max}\ll 2r$ closes the residual quantitative gap,
    confirming that macroscopic transport is governed by a classical backbone dressed by short-range coherent corrections. }
    \label{var}
\end{figure}
To compute these cumulants, we start from the {diagonal approximation} [Eq.~(\ref{convolution})], whose convolution structure implies that the cumulants accumulate stepwise.
By decomposing the trajectory into
$\mathcal{R}$
independent restart cycles and a residual segment $\tilde{n}$,
we obtain
\begin{equation}
\label{eq:scaling_laws}
(\mu_n)_{\rm inc} = 
\mathcal{R} \mu_r + \mu_{\tilde{n}},
\quad
( \sigma_n^2 )_{\rm inc} = 
\mathcal{R} \sigma_r^2 + \sigma_{\tilde{n}}^2.
\end{equation} 
For $n \gg r$,
Eq.~(\ref{eq:scaling_laws}) confirms the emergence of drifted diffusion, 
with $\sigma^2_n \approx (n/r)\sigma^2_r \propto n$, the hallmark of diffusive broadening.

However, the total transport characterized by MSD depends critically on the coin symmetry.
In the symmetric limit ($\theta=\phi=\pi/2$),
the drift vanishes ($\mu_n=0$), causing the MSD to track the variance purely:
$\langle x^2(n) \rangle \approx \sigma^2_n \propto n$ [Fig.~\ref{var}(a)].
Conversely,
for general biased coins ($\mu \neq 0$),
the MSD is dominated by $\mu_n^2$,
namely $\langle x^2(n) \rangle \propto n^2$,
exhibiting ballistic scaling despite the underlying diffusive spreading [Fig.~\ref{var}(b-d)].
We stress that the statistical quantities are explicitly tunable via the restart period $r$ and coin parameters.
This is in stark contrast to decoherence-induced diffusion \cite{Brun2003,KendonTregenna2003} or random-coin walks \cite{Joye2011}, where these coefficients are not independently controllable.
Inside each interval ($k \le r$), coherence builds up ballistically:
$\mu_k \approx \mathcal{A}(\theta, \phi) \xi \,k \propto k$ 
and $\sigma^2_k \approx \xi [ 1 - \xi \mathcal{A}^2(\theta, \phi) ] \,k^2 
\propto k^2$,
with $\xi = 1 - 1/\sqrt{2}$
and $\mathcal{A} = \cos{\theta} +\sin{\theta}\cos{\phi}$
\cite{konno2002quantum}. 
Consequently,
the drift velocity
$v_{d_{\rm inc}} = \mu_r/r \sim O(1)$
and diffusivity
$D_{\rm inc} = \sigma_r^2/2r \sim r$
can be controlled by varying the reset timing.

The interplay between ballistic cycles and diffusive accumulation
manifests as a unique staircase modulation in the statistical quantities.
As seen in Fig.~\ref{var}, the incoherent prediction (blue curves) captures the periodic fluctuations arising from the residual term $\sigma^2_{\tilde{n}}$,
confirming the survival of short-time coherent features within the diffusive envelope.

Despite this qualitative agreement,
a quantitative gap persists:
the diagonal ansatz systematically underestimates the true growth rate (see blue vs. orange in Fig.~\ref{var}).
This indicates that
the drift $v_{d_{\rm inc}}$
and diffusivity $D_{\rm inc}$
serve only as bare zeroth-order coefficients.
The missing contribution, essential for the precise agreement seen in the black curves,
arises from the quantum interference between reset-allocated sources,
that is not captured by Eq. \eqref{convolution}.
We will address this in the following section.

\begin{figure}
    \centering
    \includegraphics[width=1.0\linewidth]{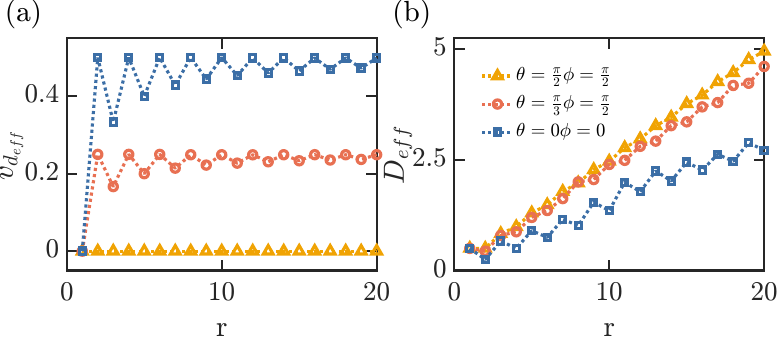}
    \caption{Long-time effective transport coefficients versus restart period $r$,  
    defined as 
    $v_{d_{\mathrm{eff}}} = [\mu_{{\cal R}r}-\mu_{({\cal R}-1)r}]/r$ 
    and
    $D_{\mathrm{eff}} = [\sigma^{2}_{{\cal R}r}-\sigma^{2}_{({\cal R}-1)r}]/2r$,
    for three coin states: 
    $(\theta,\phi)=(\pi/2,\pi/2)$ (yellow triangles), 
    $(\pi/3,\pi/2)$ (red circles), 
    and $(0,0)$ (blue squares). 
    At $r=1$, the dynamics reduces to a classical random walk; for $r > 1$, 
    $v_{d_{\mathrm{eff}}}$ saturates to coin-dependent values,
    while $D_{\mathrm{eff}}$ grows linearly with $r$.
    Thus subspace restart engineers emergent drifted diffusion in DTQWs.}
    \label{effective drift}
\end{figure}

{\em Effective light cones.}---
We argue that each reset-allocated source carries an effective light cone,
a finite spatial radius
$\Delta_{\max}\ll 2r$ 
within which its emitted wavelets remain phase-coherent and interfere constructively with those of neighboring sources.
The localized coherence is what dresses the bare diagonal coefficients 
into the renormalized transport observed in Fig.~\ref{effective drift}.
Two limits anchor the picture:
a collapsed cone $\Delta_{\max}=0$ recovers the {diagonal approximation}, 
while $\Delta_{\max}\to 2r$ restores the exact dynamics.
Our central claim is that the system sits firmly in the
small-$\Delta_{\max}$ corner of this hierarchy.

To make this quantitative,
we decompose the exact distribution in Eq. \eqref{iteration} at time $n = r\mathcal{R} + \tilde{n}$ ($\mathcal{R}\ge 0$, $0 < \tilde{n} \le r$) according to the signed spatial separation 
$\Delta = x^\prime - x^{\prime \prime}$ 
between pairs of reset-allocated sources.
Imposing a finite cutoff $\Delta_{\max}$ defines a truncated distribution that stepwise reintroduces inter-source coherence,
\begin{equation}\label{truncation}
        P^{\rm tru}_{r}(x,n) \approx 
                \sum_{\Delta = -\Delta_{\max}}^{\Delta_{\max}} 
                {\cal C}_\Delta(x,n),
\end{equation}
where $\mathcal{C}_{\Delta}(x,n) = \sum_{x^\prime} w_{\Delta}(x^\prime, r{\cal R})\, {\cal I}_{\Delta}(x-x^\prime, \tilde{n})$ isolates the probability modification from coupling sources separated by $\Delta$.
The coherence weight 
$w_\Delta(x^\prime, \mathcal{R}r) 
= \sqrt{P_{r}(x^\prime, \mathcal{R}r)\, 
P_{r}(x^\prime{-}\Delta, \mathcal{R}r)}$
is fixed by the source distribution at the last reset,
while the dynamical correlation 
${\cal I}_{\Delta}(j,\tilde{n}) = \sum_{s} \alpha_s^*(j{+}\Delta,\tilde{n}|0) \,
\alpha_s(j,\tilde{n}|0)$ 
measures the phase overlap between two wavelets emitted from sources separated by $\Delta$ at the last reset and have since propagated for $\tilde n$ steps.

The truncated $k$-th moment, with Eq.~\eqref{truncation}, is given by
$\langle x^k(n) \rangle_{\rm tru} \approx 
\sum_{ \Delta=-\Delta_{\max}}^{\Delta_{\max}} 
\sum_x   
x^k \,\mathcal{C}_{\Delta}(x,n)$.
These truncated moments reproduce with high accuracy the exact dynamics already for $\Delta_{\max}$
well below the kinematic bound $2r$ set by the maximum overlap of two ballistic wavelets,
as depicted in Fig.~\ref{var}.
The contraction reflects the interplay between the subspace reset and the intrinsic Hadamard dispersion: 
each reset disrupts the build-up of coin--position entanglement that would otherwise extend the coherent reach of the wavelets across the full ballistic span \cite{Carneiro2005}. 
The light cone is therefore not kinematic but effective, 
set by how far quantum correlations can propagate before the next reset truncates them.

Crucially, restoring this inter-source coherence leaves the transport two-scale (Fig.~\ref{var}, black vs.\ blue curves):
moments grow ballistically inside each reset interval, 
while on coarse scales every period of length $r$ behaves as a single effective step.
The light-cone interference terms
therefore renormalize the bare drift velocity and diffusivity 
($v_{d_\mathrm{inc}}, D_{\mathrm{inc}}$) 
into dressed values 
($v_{d_{\mathrm{eff}}}, D_{\mathrm{eff}}$), 
without altering the linear-in-$n$ scaling
by the {diagonal approximation}.
The exact numerics presented in Fig. \ref{effective drift} confirms this
across parameter space:
as $r$ increases, 
$v_{d_{\mathrm{eff}}}$ saturates to
coin-dependent values,
while $D_{\mathrm{eff}}$ rises linearly in $r$, 
their robustness inherited from the classical backbone and their fine structure from the finite light cone.

\textit{Discussions}.---
We have shown that periodic subspace restart engineers drifted
diffusion in quantum walks.  
The core physical picture is 
a robust {classical backbone}---the diagonal population
transport---dressed by {short-range quantum correlations}
confined within an effective light cone of radius
$\Delta_{\max}\ll 2r$. 
By curbing the build-up of coin--position entanglement, 
the restart reduces quantum interference to a local renormalization of the bare drift and diffusivity,
leaving their linear-in-$n$ scaling intact;
both coefficients remain fully tunable through the restart period and the geometric bias of the reset coin.

%{\B refs.}
The snapshot-and-reprepare character of the restart, 
namely preserving the spatial probability distribution while refreshing the coin,
maps naturally onto the active feed-forward control now standard on quantum-walk platforms \cite{carolan2015universal,nitsche2016quantum,goel2025quantuminformationprocessingspatially}. 
In spatial-mode photonic walks, the intensity at the reset epoch can be read out and replayed through a spatial light modulator \cite{li2020photonic,carolan2015universal};
in time-multiplexed architectures, the same disentangling operation follows from deterministic, properly timed polarization switching \cite{nitsche2016quantum,PhysRevLett.104.050502}.
Verification of our predictions is therefore within reach of existing experiments.

Our conclusions are not tied to the Hadamard coin or to a spatial lattice.
Because the mechanism requires only an internal state steering a translation-invariant propagation,
it carries over to any discrete-time quantum walk realized in a synthetic dimension,
e.g. orbital angular momentum 
\cite{Filippo2015}, 
frequency 
\cite{Zhang2026},
time bin \cite{Dhinwa2025},
or Fock-state and gauge-field lattices
\cite{Pang2025,ferraro2026synthetic}.
In this setting the protocol bridges two active areas: it furnishes a stroboscopic counterpart to decoherent quantum walks \cite{Brun2003},
and a discrete-time instance of collision models
with the reset coin acting as a freshly reinitialized ancilla \cite{Ciccarello2022}.
Future work may randomize the reset intervals \cite{evans2020stochastic}, 
turning the restart schedule into a further knob on the quantum-to-classical transition.

\begin{acknowledgments}
	This work is supported by the National Natural Science Foundation of China (Grants Nos. 92265208 and 12074428) and the National Key R\&D Program (Grant No. 2022YFA1405300).
\end{acknowledgments}

\bibliography{Restart}

@book{feller1971introduction,
  title={An introduction to probability theory and its applications},
  author={Feller, William},
  year={1991},
  publisher={Wiley New York},
  url={https://www.wiley.com/en-us/shop/general-introductory-statistics/an-introduction-to-probability-theory-and-its-applications-volume-1-3rd-edition-p-9780471257080}
}

@article{Montroll1965,
    author = {Montroll, Elliott W. and Weiss, George H.},
    title = {Random Walks on Lattices. II},
    journal = {J. Math. Phys.},
    volume = {6},
    number = {2},
    pages = {167--181},
    year = {1965},
    month = {02},
    issn = {0022-2488},
    doi = {10.1063/1.1704269},
    url = {https://doi.org/10.1063/1.1704269}
}

@book{Hughes1996random,
  author = {Hughes, Barry D},
    title = {Random Walks and Random Environments},
    publisher = {Oxford University Press},
    year = {1996},
    month = {06},
    isbn = {9780198537892},
    doi = {10.1093/oso/9780198537892.001.0001},
    url = {https://doi.org/10.1093/oso/9780198537892.001.0001}
}

@book{Portugal2013,
  title     = {Quantum Walks and Search Algorithms},
  author    = {Portugal, Renato},
  year      = {2013},
  doi       = {10.1007/978-1-4614-6336-8},
  isbn      = {978-1-4614-6335-1, 978-1-4614-6336-8},
  publisher = {Springer New York},
  series    = {Quantum Science and Technology},
  url       = {https://link.springer.com/book/10.1007/978-1-4614-6336-8}
}

@article{Gutmann1998,
  title = {Quantum computation and decision trees},
  author = {Farhi, Edward and Gutmann, Sam},
  journal = {Phys. Rev. A},
  volume = {58},
  issue = {2},
  pages = {915--928},
  numpages = {0},
  year = {1998},
  month = {Aug},
  publisher = {American Physical Society},
  doi = {10.1103/PhysRevA.58.915},
  url = {https://link.aps.org/doi/10.1103/PhysRevA.58.915}
}

@article{MULKEN2011,
title = {Continuous-time quantum walks: Models for coherent transport on complex networks},
journal = {Phys. Rep.},
volume = {502},
number = {2},
pages = {37--87},
year = {2011},
issn = {0370-1573},
doi = {10.1016/j.physrep.2011.01.002},
url = {https://www.sciencedirect.com/science/article/pii/S0370157311000184},
author = {M{\"u}lken, Oliver and Blumen, Alexander}
}

@misc{nayak2000quantumwalkline,
      title={Quantum Walk on the Line}, 
      author={Ashwin Nayak and Ashvin Vishwanath},
      year={2000},
      eprint={quant-ph/0010117},
      archivePrefix={arXiv},
      primaryClass={quant-ph},
      url={https://arxiv.org/abs/quant-ph/0010117}, 
}

@article{konno2002quantum,
  title={Quantum random walks in one dimension},
  author={Konno, Norio},
  journal={Quantum Inf. Process.},
  volume={1},
  number={5},
  pages={345--354},
  year={2002},
  publisher={Springer},
  doi = {10.1023/A:1023413713008},
  url = {https://link.springer.com/article/10.1023/A:1023413713008#citeas}
}

@article{PhysRevA.48.1687,
  title = {Quantum random walks},
  author = {Aharonov, Y. and Davidovich, L. and Zagury, N.},
  journal = {Phys. Rev. A},
  volume = {48},
  issue = {2},
  pages = {1687--1690},
  numpages = {0},
  year = {1993},
  month = {Aug},
  publisher = {American Physical Society},
  doi = {10.1103/PhysRevA.48.1687},
  url = {https://link.aps.org/doi/10.1103/PhysRevA.48.1687}
}

@article{PhysRevA.77.032326,
  title = {Optimizing the discrete-time quantum walk using an {SU(2)} coin},
  author = {Chandrashekar, C. M. and Srikanth, R. and Laflamme, Raymond},
  journal = {Phys. Rev. A},
  volume = {77},
  issue = {3},
  pages = {032326},
  numpages = {5},
  year = {2008},
  month = {Mar},
  publisher = {American Physical Society},
  doi = {10.1103/PhysRevA.77.032326},
  url = {https://link.aps.org/doi/10.1103/PhysRevA.77.032326}
}

@article{evans2011diffusion,
  title = {Diffusion with Stochastic Resetting},
  author = {Evans, Martin R. and Majumdar, Satya N.},
  journal = {Phys. Rev. Lett.},
  volume = {106},
  issue = {16},
  pages = {160601},
  numpages = {4},
  year = {2011},
  month = {Apr},
  publisher = {American Physical Society},
  doi = {10.1103/PhysRevLett.106.160601},
  url = {https://link.aps.org/doi/10.1103/PhysRevLett.106.160601}
}

@article{evans2020stochastic,
  title={Stochastic resetting and applications},
  author={Evans, Martin R and Majumdar, Satya N and Schehr, Gr{\'e}gory},
  journal={J. Phys. A: Math. Theor.},
  volume={53},
  number={19},
  pages={193001},
  year={2020},
  publisher={IOP Publishing},
  doi = {10.1088/1751-8121/ab7cfe},
  url = {https://iopscience.iop.org/article/10.1088/1751-8121/ab7cfe/meta}
}

@article{pal2017first,
  title = {First Passage under Restart},
  author = {Pal, Arnab and Reuveni, Shlomi},
  journal = {Phys. Rev. Lett.},
  volume = {118},
  issue = {3},
  pages = {030603},
  numpages = {6},
  year = {2017},
  month = {Jan},
  publisher = {American Physical Society},
  doi = {10.1103/PhysRevLett.118.030603},
  url = {https://link.aps.org/doi/10.1103/PhysRevLett.118.030603}
}

@article{chechkin2018random,
  title = {Random Search with Resetting: A Unified Renewal Approach},
  author = {Chechkin, A. and Sokolov, I. M.},
  journal = {Phys. Rev. Lett.},
  volume = {121},
  issue = {5},
  pages = {050601},
  numpages = {5},
  year = {2018},
  month = {Aug},
  publisher = {American Physical Society},
  doi = {10.1103/PhysRevLett.121.050601},
  url = {https://link.aps.org/doi/10.1103/PhysRevLett.121.050601}
}

@article{pal2016diffusion,
  title={Diffusion under time-dependent resetting},
  author={Pal, Arnab and Kundu, Anupam and Evans, Martin R},
  journal={J. Phys. A: Math. Theor.},
  volume={49},
  number={22},
  pages={225001},
  year={2016},
  publisher={IOP Publishing},
  doi = {10.1088/1751-8113/49/22/225001},
  url = {https://iopscience.iop.org/article/10.1088/1751-8113/49/22/225001/meta}
}

@article{reuveni2016optimal,
  title = {Optimal Stochastic Restart Renders Fluctuations in First Passage Times Universal},
  author = {Reuveni, Shlomi},
  journal = {Phys. Rev. Lett.},
  volume = {116},
  issue = {17},
  pages = {170601},
  numpages = {6},
  year = {2016},
  month = {Apr},
  publisher = {American Physical Society},
  doi = {10.1103/PhysRevLett.116.170601},
  url = {https://link.aps.org/doi/10.1103/PhysRevLett.116.170601}
}

@article{10.1063/5.0261830,
    author = {Sandev, Trifce and Iomin, Alexander and Kurths, J{\"u}rgen and Kocarev, Ljupco},
    title = {Shear-driven anomalous diffusion: Memory effects and stochastic resetting},
    journal = {Phys. Fluids},
    volume = {37},
    number = {6},
    pages = {067101},
    year = {2025},
    month = {06},
    issn = {1070-6631},
    doi = {10.1063/5.0261830},
    url = {https://doi.org/10.1063/5.0261830}
}

@misc{biswas2026threshold,
      title={Optimal threshold resetting in collective diffusive search}, 
      author={Arup Biswas and Satya N Majumdar and Arnab Pal},
      year={2026},
      eprint={2603.25338},
      archivePrefix={arXiv},
      primaryClass={cond-mat.stat-mech},
      url={https://arxiv.org/abs/2603.25338}, 
}

@article{PhysRevLett.130.050802,
  title = {Restart Expedites Quantum Walk Hitting Times},
  author = {Yin, R. and Barkai, E.},
  journal = {Phys. Rev. Lett.},
  volume = {130},
  issue = {5},
  pages = {050802},
  numpages = {6},
  year = {2023},
  month = {Feb},
  publisher = {American Physical Society},
  doi = {10.1103/PhysRevLett.130.050802},
  url = {https://link.aps.org/doi/10.1103/PhysRevLett.130.050802}
}

@article{yin2025restart,
  title={Restart uncertainty relation for monitored quantum dynamics},
  author={Yin, Ruoyu and Wang, Qingyuan and Tornow, Sabine and Barkai, Eli},
  journal={Proc. Natl. Acad. Sci. U.S.A.},
  volume={122},
  number={1},
  pages={e2402912121},
  year={2025},
  publisher={National Academy of Sciences},
  doi = {10.1073/pnas.2402912121},
  url = {https://www.pnas.org/doi/10.1073/pnas.2402912121}
}

@article{Yin2025resonance,
    author = {Yin, Ruoyu and Wang, Qingyuan and Tornow, Sabine and Barkai, Eli},
    title = {Resonances of recurrence time of monitored quantum walks},
    journal = {J. Chem. Phys.},
    volume = {162},
    number = {24},
    pages = {244114},
    year = {2025},
    month = {06},
    issn = {0021-9606},
    doi = {10.1063/5.0265944},
    url = {https://doi.org/10.1063/5.0265944}
}

@article{PhysRevResearch.7.023069,
  title = {Accelerated first detection in discrete-time quantum walks using sharp restarts},
  author = {Shukla, Kunal and Chatterjee, Riddhi and Chandrashekar, C. M.},
  journal = {Phys. Rev. Res.},
  volume = {7},
  issue = {2},
  pages = {023069},
  numpages = {8},
  year = {2025},
  month = {Apr},
  publisher = {American Physical Society},
  doi = {10.1103/PhysRevResearch.7.023069},
  url = {https://link.aps.org/doi/10.1103/PhysRevResearch.7.023069}
}

@article{PhysRevE.110.034132,
  title = {Quest for optimal quantum resetting: Protocols for a particle on a chain},
  author = {Chatterjee, Pallabi and Aravinda, S. and Modak, Ranjan},
  journal = {Phys. Rev. E},
  volume = {110},
  issue = {3},
  pages = {034132},
  numpages = {14},
  year = {2024},
  month = {Sep},
  publisher = {American Physical Society},
  doi = {10.1103/PhysRevE.110.034132},
  url = {https://link.aps.org/doi/10.1103/PhysRevE.110.034132}
}

@article{PhysRevB.98.104309,
  title = {Quantum dynamics with stochastic reset},
  author = {Mukherjee, B. and Sengupta, K. and Majumdar, Satya N.},
  journal = {Phys. Rev. B},
  volume = {98},
  issue = {10},
  pages = {104309},
  numpages = {14},
  year = {2018},
  month = {Sep},
  publisher = {American Physical Society},
  doi = {10.1103/PhysRevB.98.104309},
  url = {https://link.aps.org/doi/10.1103/PhysRevB.98.104309}
}

@article{PhysRevB.104.L180302,
  title = {Designing nonequilibrium states of quantum matter through stochastic resetting},
  author = {Perfetto, Gabriele and Carollo, Federico and Magoni, Matteo and Lesanovsky, Igor},
  journal = {Phys. Rev. B},
  volume = {104},
  issue = {18},
  pages = {L180302},
  numpages = {6},
  year = {2021},
  month = {Nov},
  publisher = {American Physical Society},
  doi = {10.1103/PhysRevB.104.L180302},
  url = {https://link.aps.org/doi/10.1103/PhysRevB.104.L180302}
}

@book{Redner_2001, 
 place={Cambridge}, 
 title={A Guide to First-Passage Processes}, 
 publisher={Cambridge University Press}, 
 author={Redner, Sidney}, 
 year={2001},
 url={https://www.cambridge.org/core/books/guide-to-firstpassage-processes/59066FD9754B42D22B028E33726D1F07}
}

@article{Majumdar2023entanglement,
  title = {Generating entanglement by quantum resetting},
  author = {Kulkarni, Manas and Majumdar, Satya N.},
  journal = {Phys. Rev. A},
  volume = {108},
  issue = {6},
  pages = {062210},
  numpages = {13},
  year = {2023},
  month = {Dec},
  publisher = {American Physical Society},
  doi = {10.1103/PhysRevA.108.062210},
  url = {https://link.aps.org/doi/10.1103/PhysRevA.108.062210}
}

@article{li2020photonic,
  title={Photonic realization of quantum resetting},
  author={Li, Zheng-Da and Yin, Xu-Fei and Wang, Zizhu and Liu, Li-Zheng and Zhang, Rui and Zhang, Yu-Zhe and Jiang, Xiao and Zhang, Jun and Li, Li and Liu, Nai-Le and others},
  journal={Optica},
  volume={7},
  number={7},
  pages={766--770},
  year={2020},
  publisher={Optical Society of America},
  doi = {10.1364/OPTICA.389322},
  url = {https://opg.optica.org/optica/fulltext.cfm?uri=optica-7-7-766}
}

@article{wald2025stochastic,
  title={Stochastic resetting in discrete-time quantum dynamics: steady states and correlations in few-qubit systems},
  author={Wald, Sascha and Yao, Louie Hong and Platini, Thierry and Hooley, Chris and Carollo, Federico},
  journal={Quantum},
  volume={9},
  pages={1742},
  year={2025},
  publisher={Verein zur F{\"o}rderung des Open Access Publizierens in den Quantenwissenschaften},
  doi = {10.22331/q-2025-05-13-1742},
  url = {https://doi.org/10.22331/q-2025-05-13-1742}
}

@article{PhysRevE.111.044143,
  title = {Discrete-time walk on one-dimensional lattice under stochastic resetting: Advantage of quantum over classical scenario},
  author = {Che\l{}miniak, Przemys\l{}aw and W\'ojcik, Jan and W\'ojcik, Antoni},
  journal = {Phys. Rev. E},
  volume = {111},
  issue = {4},
  pages = {044143},
  numpages = {11},
  year = {2025},
  month = {Apr},
  publisher = {American Physical Society},
  doi = {10.1103/PhysRevE.111.044143},
  url = {https://link.aps.org/doi/10.1103/PhysRevE.111.044143}
}

@article{Liu2023Semi,
  title = {{Semi-Markov} processes in open quantum systems. II. Counting statistics with resetting},
  author = {Liu, Fei},
  journal = {Phys. Rev. E},
  volume = {108},
  issue = {6},
  pages = {064101},
  numpages = {12},
  year = {2023},
  month = {Dec},
  publisher = {American Physical Society},
  doi = {10.1103/PhysRevE.108.064101},
  url = {https://link.aps.org/doi/10.1103/PhysRevE.108.064101}
}

@article{rbtb-8d27,
  title = {Causality, localization, and universality of monitored quantum walks with long-range hopping},
  author = {Roy, Sayan and Gupta, Shamik and Morigi, Giovanna},
  journal = {Phys. Rev. E},
  volume = {112},
  issue = {4},
  pages = {044146},
  numpages = {17},
  year = {2025},
  month = {Oct},
  publisher = {American Physical Society},
  doi = {10.1103/rbtb-8d27},
  url = {https://link.aps.org/doi/10.1103/rbtb-8d27}
}

@article{Puente2024quantumstate,
  doi = {10.22331/q-2024-03-27-1299},
  url = {https://doi.org/10.22331/q-2024-03-27-1299},
  title = {Quantum state preparation via engineered ancilla resetting},
  author = {Puente, Daniel Alcalde and Motzoi, Felix and Calarco, Tommaso and Morigi, Giovanna and Rizzi, Matteo},
  journal = {{Quantum}},
  issn = {2521-327X},
  publisher = {{Verein zur F{\"{o}}rderung des Open Access Publizierens in den Quantenwissenschaften}},
  volume = {8},
  pages = {1299},
  month = mar,
  year = {2024}
}

@article{Kumar2025,
doi = {10.1088/1751-8121/ae0bcc},
url = {https://doi.org/10.1088/1751-8121/ae0bcc},
year = {2025},
month = {oct},
publisher = {IOP Publishing},
volume = {58},
number = {41},
pages = {415002},
author = {Kumar, Ashutosh and Lahiri, Sourabh and Bagarti, Trilochan and Banerjee, Subhashish},
title = {Two and three-state quantum heat engines with stochastic resetting},
journal = {J. Phys. A: Math. Theor.}
}

@article{Bao2025,
  title = {Accelerating Quantum Relaxation via Temporary Reset: A {Mpemba}-Inspired Approach},
  author = {Bao, Ruicheng and Hou, Zhonghuai},
  journal = {Phys. Rev. Lett.},
  volume = {135},
  issue = {15},
  pages = {150403},
  numpages = {10},
  year = {2025},
  month = {Oct},
  publisher = {American Physical Society},
  doi = {10.1103/g94p-7421},
  url = {https://link.aps.org/doi/10.1103/g94p-7421}
}

@misc{solanki2025relaxation,
      title={Universal relaxation speedup in open quantum systems through transient conditional and unconditional resetting}, 
      author={Parvinder Solanki and Igor Lesanovsky and Gabriele Perfetto},
      year={2025},
      eprint={2512.10005},
      archivePrefix={arXiv},
      primaryClass={cond-mat.stat-mech},
      url={https://arxiv.org/abs/2512.10005}, 
}

@misc{gotta2026scars,
      title={Towers of quantum many-body scars under stochastic resetting}, 
      author={Lorenzo Gotta and Manas Kulkarni and Gabriele Perfetto},
      year={2026},
      eprint={2603.13165},
      archivePrefix={arXiv},
      primaryClass={cond-mat.stat-mech},
      url={https://arxiv.org/abs/2603.13165}, 
}

@misc{murauer2026,
      title={Nonequilibrium steady states induced by stochastic mid-circuit measurements and resets on a quantum computer}, 
      author={Jakob Murauer and Sabine Tornow and Gabriele Perfetto},
      year={2026},
      eprint={2606.19027},
      archivePrefix={arXiv},
      primaryClass={quant-ph},
      url={https://arxiv.org/abs/2606.19027}, 
}

@misc{carollo2026nonMarkovianity,
      title={Stochastic resetting induces quantum non-Markovianity}, 
      author={Federico Carollo and Sascha Wald},
      year={2026},
      eprint={2601.13367},
      archivePrefix={arXiv},
      primaryClass={quant-ph},
      url={https://arxiv.org/abs/2601.13367}, 
}

@article{Jafari2026,
  title = {Topological and nontopological defects from quantum reset dynamics},
  author = {Jafari, R. and Johannesson, Henrik and Eggert, Sebastian},
  journal = {Phys. Rev. B},
  volume = {113},
  issue = {22},
  pages = {224312},
  numpages = {8},
  year = {2026},
  month = {Jun},
  publisher = {American Physical Society},
  doi = {10.1103/qfdf-qw9m},
  url = {https://link.aps.org/doi/10.1103/qfdf-qw9m}
}

@article{Brun2003,
  title = {Quantum random walks with decoherent coins},
  author = {Brun, Todd A. and Carteret, H. A. and Ambainis, Andris},
  journal = {Phys. Rev. A},
  volume = {67},
  issue = {3},
  pages = {032304},
  numpages = {9},
  year = {2003},
  month = {Mar},
  publisher = {American Physical Society},
  doi = {10.1103/PhysRevA.67.032304},
  url = {https://link.aps.org/doi/10.1103/PhysRevA.67.032304}
}

@article{Maloyer2007,
doi = {10.1088/1367-2630/9/4/087},
url = {https://doi.org/10.1088/1367-2630/9/4/087},
year = {2007},
month = {apr},
publisher = {},
volume = {9},
number = {4},
pages = {87},
author = {Maloyer, Olivier and Kendon, Viv},
title = {Decoherence versus entanglement in coined quantum walks},
journal = {New J. Phys.}
}

@article{Kendon2007,
author = {Kendon, Viv},
title = {Decoherence in quantum walks---a review},
year = {2007},
issue_date = {December 2007},
publisher = {Cambridge University Press},
address = {USA},
volume = {17},
number = {6},
issn = {0960-1295},
url = {https://doi.org/10.1017/S0960129507006354},
doi = {10.1017/S0960129507006354},
journal = {Math. Struct. Comput. Sci.},
month = dec,
pages = {1169--1220},
numpages = {52}
}

@article{Carneiro2005,
doi = {10.1088/1367-2630/7/1/156},
url = {https://doi.org/10.1088/1367-2630/7/1/156},
year = {2005},
month = {jul},
publisher = {},
volume = {7},
number = {1},
pages = {156},
author = {Carneiro, Ivens and Loo, Meng and Xu, Xibai and Girerd, Mathieu and Kendon, Viv and Knight, Peter L},
title = {Entanglement in coined quantum walks on regular graphs},
journal = {New J. Phys.}
}

@article{Ciccarello2022,
title = {Quantum collision models: Open system dynamics from repeated interactions},
journal = {Phys. Rep.},
volume = {954},
pages = {1--70},
year = {2022},
issn = {0370-1573},
doi = {10.1016/j.physrep.2022.01.001},
url = {https://www.sciencedirect.com/science/article/pii/S0370157322000035},
author = {Francesco Ciccarello and Salvatore Lorenzo and Vittorio Giovannetti and G. Massimo Palma}
}

@article{Gupta2024,
  title = {{Kuramoto} model subject to subsystem resetting: How resetting a part of the system may synchronize the whole of it},
  author = {Majumder, Rupak and Chattopadhyay, Rohitashwa and Gupta, Shamik},
  journal = {Phys. Rev. E},
  volume = {109},
  issue = {6},
  pages = {064137},
  numpages = {22},
  year = {2024},
  month = {Jun},
  publisher = {American Physical Society},
  doi = {10.1103/PhysRevE.109.064137},
  url = {https://link.aps.org/doi/10.1103/PhysRevE.109.064137}
}

@article{Gupta2025,
  title = {Manipulating Phases in Many-Body Interacting Systems with Subsystem Resetting},
  author = {Acharya, Anish and Majumder, Rupak and Gupta, Shamik},
  journal = {Phys. Rev. Lett.},
  volume = {135},
  issue = {12},
  pages = {127103},
  numpages = {10},
  year = {2025},
  month = {Sep},
  publisher = {American Physical Society},
  doi = {10.1103/np7q-hxld},
  url = {https://link.aps.org/doi/10.1103/np7q-hxld}
}

@misc{majumder2026analytical,
  title         = {Analytical approach to subsystem resetting in generalized {Kuramoto} models},
  author        = {Majumder, Rupak and Acharya, Anish and Gupta, Shamik},
  year          = {2026},
  eprint        = {2604.04769},
  archivePrefix = {arXiv},
  primaryClass  = {cond-mat.stat-mech},
  url           = {https://arxiv.org/abs/2604.04769}
}

@article{carolan2015universal,
  title={Universal linear optics},
  author={Carolan, Jacques and Harrold, Christopher and Sparrow, Chris and Mart{\'\i}n-L{\'o}pez, Enrique and Russell, Nicholas J and Silverstone, Joshua W and Shadbolt, Peter J and Matsuda, Nobuyuki and Oguma, Manabu and Itoh, Mikitaka and others},
  journal={Science},
  volume={349},
  number={6249},
  pages={711--716},
  year={2015},
  publisher={American Association for the Advancement of Science},
  doi = {10.1126/science.aab3642},
  url = {https://www.science.org/doi/10.1126/science.aab3642},
}

@article{nitsche2016quantum,
  title={Quantum walks with dynamical control: graph engineering, initial state preparation and state transfer},
  author={Nitsche, Thomas and Elster, Fabian and Novotn{\'y}, Jaroslav and G{\'a}bris, Aur{\'e}l and Jex, Igor and Barkhofen, Sonja and Silberhorn, Christine},
  journal={New J. Phys.},
  volume={18},
  number={6},
  pages={063017},
  year={2016},
  publisher={IOP Publishing},
  doi = {10.1088/1367-2630/18/6/063017},
  url = {https://iopscience.iop.org/article/10.1088/1367-2630/18/6/063017}
}

@article{PhysRevLett.104.050502,
  title = {Photons Walking the Line: A Quantum Walk with Adjustable Coin Operations},
  author = {Schreiber, A. and Cassemiro, K. N. and Poto{\v{c}}ek, V. and G{\'a}bris, A. and Mosley, P. J. and Andersson, E. and Jex, I. and Silberhorn, Ch.},
  journal = {Phys. Rev. Lett.},
  volume = {104},
  issue = {5},
  pages = {050502},
  numpages = {4},
  year = {2010},
  month = {Feb},
  publisher = {American Physical Society},
  doi = {10.1103/PhysRevLett.104.050502},
  url = {https://link.aps.org/doi/10.1103/PhysRevLett.104.050502}
}

@misc{goel2025quantuminformationprocessingspatially,
      title={Quantum Information Processing with Spatially Structured Light}, 
      author={Suraj Goel and Bohnishikha Ghosh and Mehul Malik},
      year={2025},
      eprint={2510.11154},
      archivePrefix={arXiv},
      primaryClass={quant-ph},
      url={https://arxiv.org/abs/2510.11154}, 
}

@article{Kempe2003,
  author  = {Kempe, Julia},
  title   = {Quantum random walks: An introductory overview},
  journal = {Contemp. Phys.},
  volume  = {44},
  number  = {4},
  pages   = {307--327},
  year    = {2003},
  doi     = {10.1080/00107151031000110776},
  url     = {https://doi.org/10.1080/00107151031000110776}
}

@article{VenegasAndraca2012,
  author  = {Venegas-Andraca, Salvador El{\'i}as},
  title   = {Quantum walks: a comprehensive review},
  journal = {Quantum Inf. Process.},
  volume  = {11},
  number  = {5},
  pages   = {1015--1106},
  year    = {2012},
  doi     = {10.1007/s11128-012-0432-5},
  url     = {https://doi.org/10.1007/s11128-012-0432-5}
}

@inproceedings{Ambainis2001,
  author    = {Ambainis, Andris and Bach, Eric and Nayak, Ashwin and Vishwanath, Ashvin and Watrous, John},
  title     = {One-dimensional quantum walks},
  booktitle = {Proceedings of the 33rd Annual ACM Symposium on Theory of Computing},
  pages     = {37--49},
  publisher = {Association for Computing Machinery},
  year      = {2001},
  doi       = {10.1145/380752.380757},
  url       = {https://doi.org/10.1145/380752.380757}
}

@article{Mackay2002,
  author  = {Mackay, T. D. and Bartlett, S. D. and Stephenson, L. T. and Sanders, B. C.},
  title   = {Quantum walks in higher dimensions},
  journal = {J. Phys. A: Math. Gen.},
  volume  = {35},
  number  = {12},
  pages   = {2745--2753},
  year    = {2002},
  doi     = {10.1088/0305-4470/35/12/304},
  url     = {https://doi.org/10.1088/0305-4470/35/12/304}
}

@article{Tregenna2003,
  author  = {Tregenna, Ben and Flanagan, Will and Maile, Rik and Kendon, Viv},
  title   = {Controlling discrete quantum walks: coins and initial states},
  journal = {New J. Phys.},
  volume  = {5},
  pages   = {83},
  year    = {2003},
  doi     = {10.1088/1367-2630/5/1/383},
  url     = {https://doi.org/10.1088/1367-2630/5/1/383}
}

@article{Shenvi2003,
  author  = {Shenvi, Neil and Kempe, Julia and Whaley, K. Birgitta},
  title   = {Quantum random-walk search algorithm},
  journal = {Phys. Rev. A},
  volume  = {67},
  number  = {5},
  pages   = {052307},
  numpages = {11},
  year    = {2003},
  month = {May},
  publisher = {American Physical Society},
  doi     = {10.1103/PhysRevA.67.052307},
  url ={https://link.aps.org/doi/10.1103/PhysRevA.67.052307}
}

@inproceedings{Childs2003,
  author    = {Childs, Andrew M. and Cleve, Richard and Deotto, Enrico and Farhi, Edward and Gutmann, Sam and Spielman, Daniel A.},
  title     = {Exponential algorithmic speedup by a quantum walk},
  booktitle = {Proceedings of the 35th Annual ACM Symposium on Theory of Computing},
  pages     = {59--68},
  publisher = {Association for Computing Machinery},
  year      = {2003},
  doi       = {10.1145/780542.780552},
  url       = {https://doi.org/10.1145/780542.780552}
}

@article{Ambainis2007,
  author  = {Ambainis, Andris},
  title   = {Quantum Walk Algorithm for Element Distinctness},
  journal = {SIAM J. Comput.},
  volume  = {37},
  number  = {1},
  pages   = {210--239},
  year    = {2007},
  doi     = {10.1137/S0097539705447311},
  url     = {https://doi.org/10.1137/S0097539705447311}
}

@article{Childs2009,
  author  = {Childs, Andrew M.},
  title   = {Universal Computation by Quantum Walk},
  journal = {Phys. Rev. Lett.},
  volume  = {102},
  number  = {18},
  pages   = {180501},
  numpages = {4},
  year    = {2009},
  month = {May},
  publisher = {American Physical Society},
  doi     = {10.1103/PhysRevLett.102.180501},
  url     = {https://doi.org/10.1103/PhysRevLett.102.180501}
}

@article{KendonTregenna2003,
  author  = {Kendon, Viv and Tregenna, Ben},
  title   = {Decoherence can be useful in quantum walks},
  journal = {Phys. Rev. A},
  volume  = {67},
  number  = {4},
  pages   = {042315},
  numpages = {6},
  year    = {2003},
  month = {Apr},
  publisher = {American Physical Society},
  doi     = {10.1103/PhysRevA.67.042315},
  url     = {https://doi.org/10.1103/PhysRevA.67.042315}
}

@article{ROMANELLI2005,
title = {Decoherence in the quantum walk on the line},
journal = {Physica A (Amsterdam)},
volume = {347},
pages = {137--152},
year = {2005},
issn = {0378-4371},
doi = {10.1016/j.physa.2004.08.070},
url = {https://www.sciencedirect.com/science/article/pii/S0378437104011422},
author = {A. Romanelli and R. Siri and G. Abal and A. Auyuanet and R. Donangelo}
}

@article{Oliveira2006,
  title = {Decoherence in two-dimensional quantum walks},
  author = {Oliveira, A. C. and Portugal, R. and Donangelo, R.},
  journal = {Phys. Rev. A},
  volume = {74},
  issue = {1},
  pages = {012312},
  numpages = {8},
  year = {2006},
  month = {Jul},
  publisher = {American Physical Society},
  doi = {10.1103/PhysRevA.74.012312},
  url = {https://link.aps.org/doi/10.1103/PhysRevA.74.012312}
}

@article{Annabestani2010,
  author  = {Annabestani, Mostafa and Akhtarshenas, Seyed Javad and Abolhassani, Mohamad Reza},
  title   = {Decoherence in a one-dimensional quantum walk},
  journal = {Phys. Rev. A},
  volume  = {81},
  number  = {3},
  pages   = {032321},
  numpages = {9},
  year    = {2010},
  month = {Mar},
  publisher = {American Physical Society},
  doi     = {10.1103/PhysRevA.81.032321},
  url     = {https://doi.org/10.1103/PhysRevA.81.032321}
}

@article{Broome2010,
  author  = {Broome, M. A. and Fedrizzi, A. and Lanyon, B. P. and Kassal, I. and Aspuru-Guzik, A. and White, A. G.},
  title   = {Discrete Single-Photon Quantum Walks with Tunable Decoherence},
  journal = {Phys. Rev. Lett.},
  volume  = {104},
  number  = {15},
  pages   = {153602},
  numpages = {4},
  year    = {2010},
  month = {Apr},
  publisher = {American Physical Society},
  doi     = {10.1103/PhysRevLett.104.153602},
  url     = {https://doi.org/10.1103/PhysRevLett.104.153602}
}

@article{Schreiber2011,
  author  = {Schreiber, A. and Cassemiro, K. N. and Poto{\v{c}}ek, V. and G{\'a}bris, A. and Jex, I. and Silberhorn, Ch.},
  title   = {Decoherence and Disorder in Quantum Walks: From Ballistic Spread to Localization},
  journal = {Phys. Rev. Lett.},
  volume  = {106},
  number  = {18},
  pages   = {180403},
  numpages = {4},
  year    = {2011},
  month = {May},
  publisher = {American Physical Society},
  doi     = {10.1103/PhysRevLett.106.180403},
  url     = {https://doi.org/10.1103/PhysRevLett.106.180403}
}

@article{Joye2011,
  author  = {Joye, Alain},
  title   = {Random Time-Dependent Quantum Walks},
  journal = {Commun. Math. Phys.},
  volume  = {307},
  number  = {1},
  pages   = {65--100},
  year    = {2011},
  doi     = {10.1007/s00220-011-1297-7},
}

@article{Alberti2014,
  author  = {Alberti, Andrea and Alt, Wolfgang and Werner, Reinhard and Meschede, Dieter},
  title   = {Decoherence models for discrete-time quantum walks and their application to neutral atom experiments},
  journal = {New J. Phys.},
  volume  = {16},
  number  = {12},
  pages   = {123052},
  year    = {2014},
  doi     = {10.1088/1367-2630/16/12/123052},
  url     = {https://doi.org/10.1088/1367-2630/16/12/123052}
}

@Article{Eliahu2021,
AUTHOR = {Jayakody, Mahesh N. and Nanayakkara, Asiri and Cohen, Eliahu},
TITLE = {Analysis of Decoherence in Linear and Cyclic Quantum Walks},
JOURNAL = {Optics},
VOLUME = {2},
YEAR = {2021},
NUMBER = {4},
PAGES = {236--250},
URL = {https://www.mdpi.com/2673-3269/2/4/22},
ISSN = {2673-3269},
DOI = {10.3390/opt2040022}
}

@article{Bressanini2022,
  title     = {Decoherence and classicalization of continuous-time quantum walks on graphs},
  author    = {Bressanini, Gabriele and Benedetti, Claudia and Paris, Matteo G. A.},
  journal   = {Quantum Inf. Process.},
  volume    = {21},
  number    = {9},
  pages     = {317},
  year      = {2022},
  month     = {sep},
  publisher = {Springer},
  doi       = {10.1007/s11128-022-03647-x},
  url       = {https://doi.org/10.1007/s11128-022-03647-x}
}

@article{Kua2025,
doi = {10.1088/1402-4896/ae0c6b},
url = {https://doi.org/10.1088/1402-4896/ae0c6b},
year = {2025},
month = {oct},
publisher = {IOP Publishing},
volume = {100},
number = {10},
pages = {105108},
author = {Kua, K H and Muniandy, S V and Ishak, Nur Izzati and Chong, W Y},
title = {{Wigner} function analysis of the decoherence of discrete-time quantum walk in the bit-phase noise channel},
journal = {Phys. Scr.}
}

@article{Perets2008,
  author  = {Perets, Hagai B. and Lahini, Yoav and Pozzi, Francesca and Sorel, Marc and Morandotti, Roberto and Silberberg, Yaron},
  title   = {Realization of Quantum Walks with Negligible Decoherence in Waveguide Lattices},
  journal = {Phys. Rev. Lett.},
  volume  = {100},
  issue = {17},
  pages   = {170506},
  numpages = {4},
  year    = {2008},
  month = {May},
  publisher = {American Physical Society},
  doi     = {10.1103/PhysRevLett.100.170506},
  url     = {https://doi.org/10.1103/PhysRevLett.100.170506}
}

\newpage

\end{document}